\documentclass[pra,aps,twocolumn,showpacs,floatfix,superscriptaddress]{revtex4}
\usepackage{eurosym}
\usepackage{amssymb}
\usepackage{graphicx}
\usepackage{dcolumn}
\usepackage{bm,amsmath,verbatim}
\usepackage{hyperref}
\usepackage{epstopdf}
\usepackage{amssymb}
\usepackage{hyperref}
\usepackage{graphicx}
\usepackage{dcolumn}
\usepackage{bm,amsmath,verbatim}
\usepackage{mathrsfs}
\usepackage{mciteplus}

\def\be{\begin{equation}}
\def\ee{\end{equation}}
\def\bea{\begin{eqnarray}}
\def\eea{\end{eqnarray}}
\def\bse{\begin{subequations}}
\def\ese{\end{subequations}}

\def\be{\begin{eqnarray}}
\def\ee{\end{eqnarray}}

\begin{document}

\title{Spin-Orbit Driven Transitions Between Mott Insulators and Finite
Momentum Superfluids of Bosons in Optical Lattices}
\author{Mi Yan}
\affiliation{Department of Physics, Virginia Tech, Blacksburg, Virginia 24061 USA}
\author{Yinyin Qian}
\affiliation{Department of Physics, The University of Texas at Dallas, Richardson, Texas
75080 USA}
\author{Hoi-Yin Hui}
\affiliation{Department of Physics, Virginia Tech, Blacksburg, Virginia 24061 USA}
\author{Ming Gong}
\affiliation{Key Lab of Quantum Information, CAS, University of Science and Technology of
China, Hefei, 230026, P.R. China}
\affiliation{Synergetic Innovation Center of Quantum Information and Quantum Physics,
University of Science and Technology of China, Hefei, 230026, P.R. China}
\affiliation{Department of Physics, The University of Texas at Dallas, Richardson, Texas
75080 USA}
\author{Chuanwei Zhang}
\email{chuanwei.zhang@utdallas.edu}
\affiliation{Department of Physics, The University of Texas at Dallas, Richardson, Texas
75080 USA}
\author{V.W. Scarola}
\affiliation{Department of Physics, Virginia Tech, Blacksburg, Virginia 24061 USA}

\begin{abstract}
Synthetic spin-orbit coupling in ultracold atomic gases can be taken to
extremes rarely found in solids. We study a two dimensional Hubbard model of
bosons in an optical lattice in the presence of spin-orbit coupling strong
enough to drive direct transitions from Mott insulators to superfluids. Here
we find phase-modulated superfluids with finite momentum that are generated
entirely by spin-orbit coupling. We investigate the rich phase patterns of
the superfluids, which may be directly probed using time-of-flight imaging
of the spin-dependent momentum distribution.
\end{abstract}

\maketitle

\section{Introduction}

The Rashba effect \cite{rashba:1960} in solids derives from the motion of an
electron in a strong electric field. As the electron moves in the presence
of a potential gradient, $\nabla V$, it experiences an effective magnetic
field in its frame of reference. The Rashba energy \cite{rashba:1960}:
\begin{equation}
(\Vec{\nabla}V\times \Vec{p})\cdot \Vec{\sigma},  \label{eq-rashba}
\end{equation}%
captures the energetics of electron spin reorientation due to the effective
magnetic field, where $\Vec{p}$ is the particle momentum and $\Vec{\sigma}$
are the Pauli matrices. The Rashba spin-orbit coupling (SOC) energy is well
known to be particularly strong at metallic surfaces \cite{AgAu, AgAu1}
(e.g., on Ag(111) or Au(111)) because here we find extremely strong
potential gradients. As a result, studies of the impact of Rashba SOC on
two-dimensional (2D) conductors have a long history \cite{winkler:2003}. But
the impact of Rashba SOC on the surface states of Mott insulators has come
under more careful scrutiny recently because of possible connections to
topological insulators \cite{hasan:2010,qi:2011}.

\begin{figure}[tbp]
\includegraphics[width=3.2in]{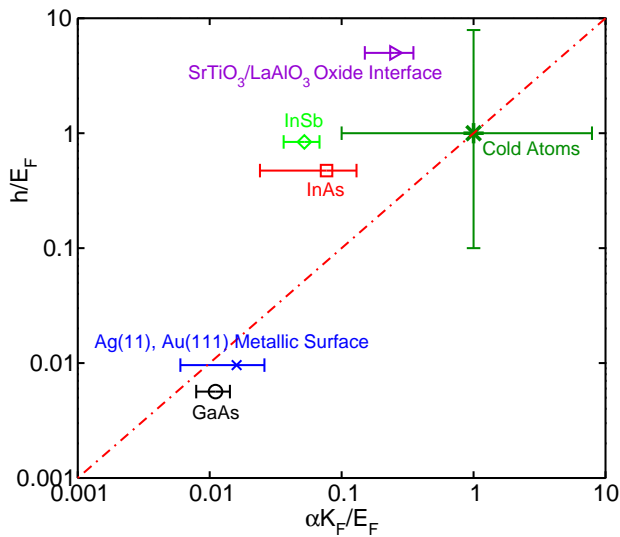}
\caption{Comparison of SOC strengths in solids and cold atoms. $h$, $\protect%
\alpha $, and $E_{F}$ denote the Zeeman energy, SOC coefficient, and Fermi
energy, respectively. For GaAs, the effective mass is $m^{\ast }=0.067m_{0}$%
\protect\cite{Band}, where $m_{0}$ is the electron mass, the Rashba SOC
strength is $\protect\alpha =(0.04-0.06)\times 10^{-11}$ eV$\cdot $m%
\protect\cite{agaas}, and the $g$-factor is $g^{\ast }=-0.45$\protect\cite%
{ggaas}. For InAs the parameters are: $m^{\ast }=0.026m_{0}$\protect\cite%
{Band}, $\protect\alpha =(0.28-1.4)\times 10^{-11}$ eV$\cdot $m\protect\cite%
{ainas}, and $g^{\ast }=-15.1$\protect\cite{Lg}. For InSb the parameters
are: $m^{\ast }=0.0135$\protect\cite{Band}, $\protect\alpha %
K_{F}=(1.0-1.2)\times 10^{-11}$ eV$\cdot $m\protect\cite{ainsb}, and $%
g^{\ast }=-51$\protect\cite{Lg}. And for the metallic surfaces: $m^{\ast
}\sim 0.255m_{0}$\protect\cite{AgAu, AgAu1}, where the $g$-factor is assumed
to be $g^{\ast }=2$. The parameters for these four different examples are
plotted at an external magnetic field of $5$ Tesla. A high carrier density, $%
n=10^{11}$ cm$^{-2}$, is used for the semiconductors. For SrTiO$_{3}$/LaAlO$%
_{3}$ oxide interfaces, the data are taken from Ref. \onlinecite{Patrick}.
Additional feasible parameter regimes are plotted as horizontal and vertical
bars. }
\label{fig-comp}\centering

\end{figure}

Mott insulators localize as a result of strong interaction and would
therefore appear to exclude the possibility of SOC effects, but one can
argue that this is not always the case. Small momentum in Eq.~(\ref%
{eq-rashba}) (the case for localized states) does not necessarily imply low
Rashba energies. In an extreme limit, Mott insulating surfaces can, in
principle, experience very large potential gradients that can compensate the
small momentum, i.e., $\langle p \rangle \rightarrow 0 $ with $\langle \Vec{%
\nabla} V\times \Vec{p} \rangle \sim E_F$, where $E_F$ is the Fermi energy.
If, in this limit, the energetics of Rashba SOC compete with the Mott gap,
one could observe a transition between a Mott insulator and a conducting
state driven entirely by Rashba SOC in spite of the small average momentum
of particles in Mott insulators. Unfortunately, the limit where Rashba SOC
competes with the Mott gap is rare in solids because it would typically be
precluded by other effects, such as charge transfer between bands. But this
limit can be explored in another context: using synthetic SOC in optical
lattices.

Recent experimental progress \cite%
{lin:2011,wang:2012,cheuk:2012,zhang:2012,fu:2013,Qu:2013,williams:2013}
demonstrates engineering of synthetic SOC for ultracold atomic gases \cite%
{bloch:2008}. These experiments show that Raman beams can be used to dress
atoms with a spin-dependent momentum. Rashba (and/or Dresselhaus) SOCs
governing these dressed states \cite{sau:2011,galitski:2013} are tunable to
extremes not possible in solids, see Fig. \ref{fig-comp}. Recent work shows,
for example, that synthetic SOC can generate flat bands \cite%
{zhang:2013,lin:2013, hui:2017, chen:2017}, exotic superfluidity \cite%
{hu:2012}, and intriguing vortex structures \cite%
{sau:2011,ramachandhran:2012,zhou:2013}.

Recent theory work has also explored the impact of SOC on the spin structure
of Mott insulators in optical lattices \cite%
{Radic12,Cole12,Cai12,Mandal,Gong12}. Here super-exchange coupling between
sites was shown to combine with Rashba SOC to lead to rich spin structures
within the Mott state \cite{Radic12,Cole12,Cai12,Mandal,Gong12}. But in
these studies parameters were chosen to explore the impact of Rashba SOC on
the spin physics of Mott insulators while leaving the charge structure
intact.

In this work we explore Rashba SOC that is strong enough to cause the
breakdown of charge ordering in Mott insulators. This extreme limit is of
direct relevance to optical lattice experiments with synthetic SOC. We
study, in particular, a 2D lattice model of two-component interacting bosons
in the presence of tunable Rashba coupling. We find that strong Rashba SOC
can cause the breakdown of the Mott insulating state and drive a direct
transition between the Mott insulator and a superfluid state, even in the $%
\emph{absence}$ of single particle tunneling between sites of the lattice
\cite{Mandal}. This limit is the lattice version of the limit discussed
above, $\langle p \rangle \rightarrow 0 $ with $\langle \Vec{\nabla} V\times
\Vec{p} \rangle \sim E_F$, where vanishing kinetics leaves Rashba SOC to
generate its own conducting state. For the case of lattice bosons studied
here, we find that Rashba SOC generates finite momentum superfluids. We show
that these superfluids are characterized by staggered phase patterns. We
also find distinct superfluid states with striped phase patterns that are
separated by transitions on finite lattices with periodic boundaries. We
predict that finite momentum superfluids should be observable in
time-of-flight measurements of the momentum distribution.

The paper is organized as follows: In Sec.~\ref{sec-modelsmethods} we
construct a Bose-Hubbard model of two-component atoms in the presence of
Rashba SOC. We also discuss two complimentary mean field approaches that
allow us to compute the phase diagram, transition properties, and the
momentum distribution. In Sec.~\ref{sec-resultsfinite} we present results on
finite lattice sizes. We use Gutzwiller mean field theory to show that
Rashba SOC causes the Bosonic Mott insulator to give way to finite momentum
superfluids. We also explore inter-superfluid transitions. We find that
transitions separate distinct phase patterns of finite momentum superfluids.
We demonstrate in Sec. ~\ref{sec-resultstrap} that these different finite
momentum phases can indeed be observed in experiments with a trapping
potential. In Sec.~\ref{sec-resultsinfinite} we present analytic arguments
that transitions depend critically on boundary effects, akin to effects
found in Fulde-Ferrell-Larkin-Ovchinnikov (FFLO) superconductors \cite%
{FF64,LO64,LO65,FFLO:Wu,RMP04,Mizushima05,Hu07,Koponen08,Feiguin07,Paananen09,Parish07,Tempere07,Loh10}%
. We show that analytic mean field calculations in the infinite system size
limit do not show these transitions. We summarize in Sec.~\ref{sec-summary}.

\section{Model and Methods}

\label{sec-modelsmethods}

We consider a 2D square optical lattice containing bosonic atoms with two
hyperfine levels. States with two hyperfine levels act as a pseudo-spin 1/2
state. We also assume the presence of Raman beams that couple the atomic
momentum to the spin to generate synthetic SOC \cite%
{lin:2011,wang:2012,cheuk:2012,zhang:2012,fu:2013,williams:2013}. The
interaction between alkali atoms is governed by a short-range ($s$-wave)
repulsion. For a deep optical lattice, the problem can be accurately
described in the single-band, tight-binding limit \cite{Jaksch98} where the $%
s$-wave interaction becomes an on-site Hubbard interaction and the SOC is
discretized.

To study this system we construct a Hubbard model of two-component bosons in
the presence of Rashba SOC on a square lattice. We allow the on-site Hubbard
interaction to have a spin-dependent interaction:
\begin{eqnarray}
H &=& -t\sum_{\langle ij \rangle}\Psi_i^{\dagger}\Psi_j^{\vphantom{\dagger}}
+ \frac{U}{2}\sum_{i\sigma}n_{i\sigma}(n_{i\sigma}-1) \quad  \notag \\
&+& U_{\uparrow \downarrow} \sum_{i}n_{i\uparrow}n_{i\downarrow}-
\mu\sum_{i\sigma}n_{i\sigma}  \notag \\
&+& i\lambda\sum_{\langle ij \rangle}\Psi^{\dagger}_i \Vec{e}_z \cdot (\Vec{%
\sigma} \times \Vec{d}_{ij} ) \Psi_j^{\vphantom{\dagger}}+ H.c.,
\label{eq-totalmodel}
\end{eqnarray}
where, $\Psi_i =(b_{i\uparrow},b_{i\downarrow})^T$ is a two-component
bosonic annihilation operator at the site $i$, $n_{{i\sigma}}=b_{i\sigma}^{{%
\dagger}}b_{i\sigma}^{\vphantom{\dagger}}$, $t$ is the spin-independent
nearest neighbor tunneling, $U$ ($U_{\uparrow \downarrow}$) is the on-site
interaction between bosons of the same (different) spin $\sigma$, and $\mu$
is the chemical potential. In the last term $\lambda$ is the Rashba SOC
strength, $\Vec{d}_{ij}$ is the unit vector between the neighboring sites $i$
and $j$, and $\Vec{e}_z$ is the unit vector along the $z$ direction. In the
following we use $U = 1$ to set the energy scale.

\begin{figure}[tbp]
\centering
\includegraphics[width=3in]{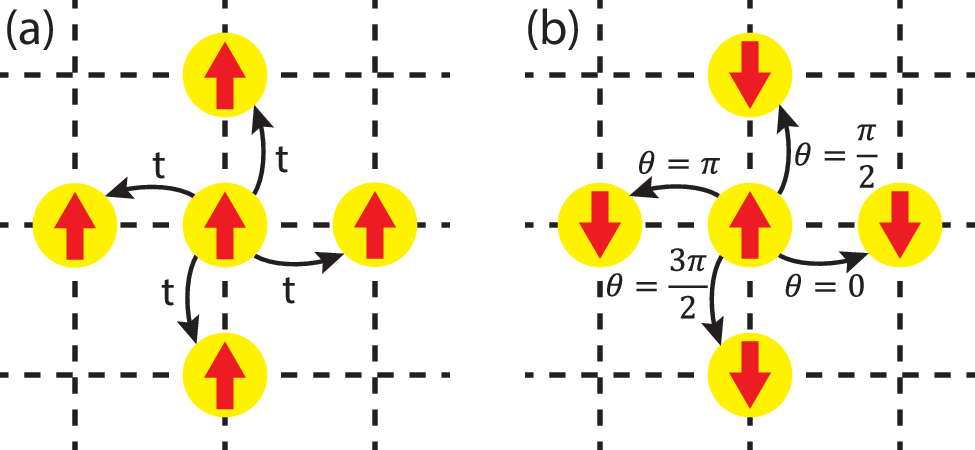}
\caption{Schematic of spin independent tunneling (a) and spin-dependent
tunneling induced by SOC (b). In the later case, the tunneling takes place
between two neighboring sites accompanied by both spin flipping and phase
variations. The phase variation during tunneling is responsible for the
creation of the finite momentum superfluids. }
\label{hoppingsfig}
\end{figure}

The tunneling and Rashba terms induce two \emph{different} types of
superfluidity. To see this we plot the spin-independent tunneling and
spin-dependent tunneling in Fig.~\ref{hoppingsfig}. The left panel shows
that the spin-independent tunneling favors phase uniformity since $t$ is
real. But in the right panel we see that SOC has two effects: It induces
tunneling between neighboring sites with two different spin states and it
imposes phase variation. The phase variation depends strongly on the
direction of the neighboring sites. SOC therefore favors highly anisotropic
superfluid states. Without SOC the system has at least an $U(1)\otimes U(1)$
symmetry, which means that the total number of each species are conserved;
however, SOC introduces spin flips between two neighboring sites, thus the
system only respects $U(1)$ symmetry and, as a result, the phase difference
between the neighboring sites can not be gauged out. The competition between
spin-independent tunneling and spin-dependent tunneling tunes the transition
between these different superfluids.

In the weakly interacting limit the model exhibits three different
superfluid phases: In the regime when spin-independent tunneling dominates ($%
t \gg \lambda$), the uniform superfluid is preferred and the total momentum
of the superfluid is zero; In the opposite regime, a staggered superfluid
phase is preferred; and in the intermediate regime, $t\sim\lambda$, the
strong competition between the two tunnelings gives rise to superfluids with
phase patterns that depend strongly on boundary effects.

Strong interactions add competing Mott insulating phases and complicates
estimates of the phase diagram. To study the competition between all ground
states we use two complimentary mean field approaches. We apply the
Gutzwiller mean field method to finite system sizes (relevant to
experiments) and compare with an otherwise equivalent mean field method
applied to infinite system sizes.

We now discuss the Gutzwiller mean field method \cite{rokhsar:1991,Jaksch98}%
. The method assumes a product ground state of the form: $|G\rangle
=\prod_{i,\sigma }\left( \sum_{n}f_{n}^{\left( i,\sigma \right) }|n\rangle
_{i,\sigma }\right) $. This form for the wavefunction has been extensively
applied to bosons in optical lattices \cite{Jaksch98}, even in the presence
of complex hopping amplitudes \cite{scarola:2007}. It generally gives
quantitatively reliable results in 2D and 3D, (for comparisons, see, e.g.,
Ref.~\cite{zakrezewski:2005}), and is a particularly excellent approximation
when computing local correlation functions (See, e.g., Ref.~\cite%
{niederle:2013}). The variational parameters $f$ are obtained by minimizing
the total energy:
\begin{equation}
E=\frac{\langle G|H|G\rangle }{\langle G|G\rangle }.
\end{equation}%
We minimize the total ground state energy with the conjugate gradient
algorithm \cite{Debye09,Deift93}. The ground state energy is reached when
the energy variation is less than $10^{-5}U$, which is sufficient to
distinguish the energy difference between different phases.

\begin{figure}[t]
\centering
\includegraphics[width=3.3in]{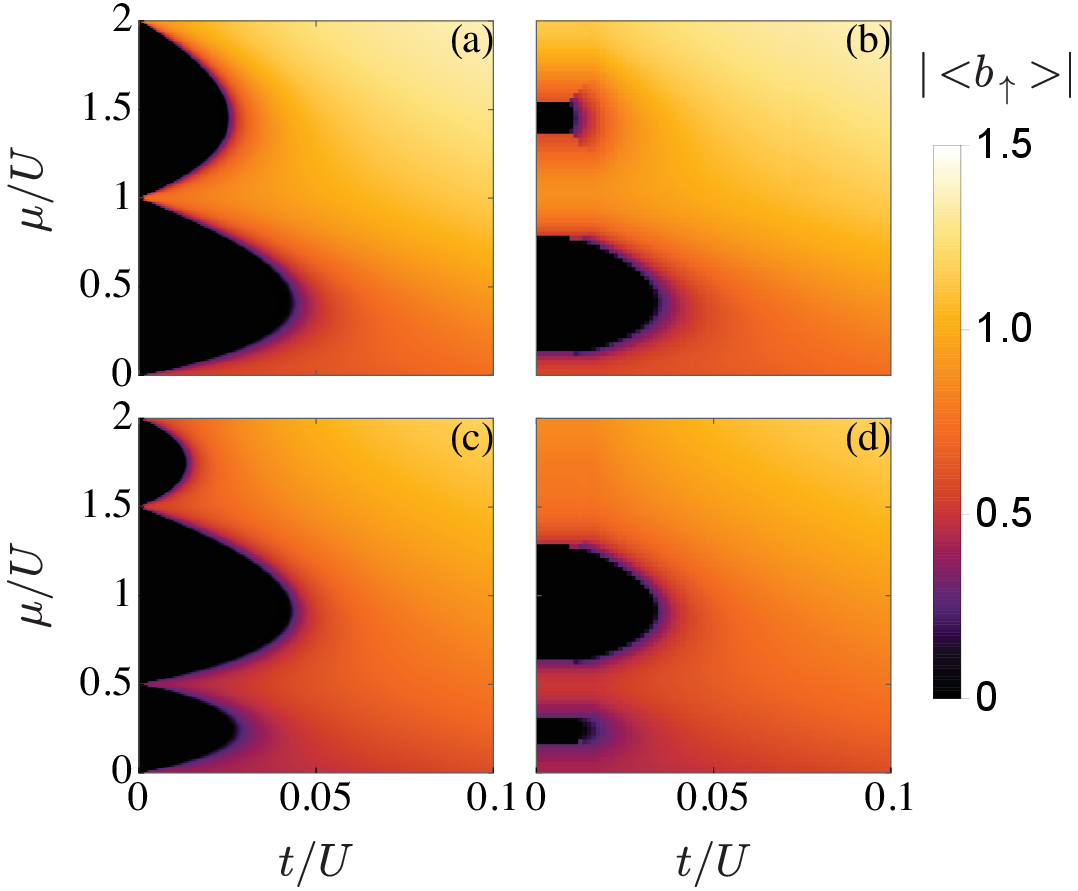}
\caption{Phase diagrams of of Eq.~\protect\ref{eq-totalmodel} obtained from
Gutzwiller variational simulations for an $8 \times 8 $ lattice with
periodic boundary condition at (a) $U_{\uparrow \downarrow}=0, \protect%
\lambda=0$, (b) $U_{\uparrow \downarrow}=0, \protect\lambda=0.04U$, (c) $%
U_{\uparrow \downarrow}=0.5U, \protect\lambda=0$, and (d) $U_{\uparrow
\downarrow}=0.5U, \protect\lambda=0.04U$. The phase diagrams are determined
by the amplitude of the spin-up superfluid order parameter. The spin-down
superfluid order parameter produces similar results. }
\label{fig-phase}
\end{figure}

We supplement the finite system size Gutzwiller method with an equivalent
mean field limit applied to infinite system sizes. We assume $\langle
b_{i\sigma }\rangle =\psi e^{i\theta _{i\sigma }}$, where $\psi $ is a real
number. This assumption is equivalent to the assumed form for $|G\rangle $
but works best on infinite system sizes. The total energy then becomes:
\begin{equation}
E_{\psi }=(U+U_{\uparrow \downarrow })\psi ^{4}-(U+2\mu +tA+\lambda B)\psi
^{2},  \label{eq-energymeanfield}
\end{equation}%
where the coefficients are:
\begin{equation}
A\equiv N^{-1}\sum_{\langle ij\rangle }\left[ e^{i(\theta _{j\uparrow
}-\theta _{i\uparrow })}+e^{i(\theta _{j\downarrow }-\theta _{i\downarrow
})}+H.c.\right] ,
\end{equation}%
and:
\begin{equation}
B\equiv N^{-1}\sum_{\langle ij\rangle }\left[ Z_{ij}^{\ast }e^{i(\theta
_{j\downarrow }-\theta _{i\uparrow })}-Z_{ij}e^{i(\theta _{j\uparrow
}-\theta _{i\downarrow })}+H.c.\right] ,
\end{equation}%
with $Z_{ij}\equiv d_{ij}^{x}+id_{ij}^{y}$ and $N$ is the number of sites.
An important point here is that the total energy depends not only on the
magnitude of the order parameter $\psi $, but also on the phase difference
between neighboring sites. We see that the minimal energy $E_{\psi }$
corresponds to a maximal value of $A$ and $B$ when $U$, $U_{\uparrow
\downarrow }$, $\lambda $, and $t$ assume positive values (the case studied
in this paper). Here $A$ depends only on the phase difference between the
same spin states, while $B$ depends strongly on the phase difference between
spin up and spin down states in the neighboring sites. The competition
between $A$ and $B$ governs competition between superfluids with distinct
phase patterns. When $\lambda =0 $, $A$ takes its maximum value when all of
the sites have the same phase, which corresponds to the uniform superfluid
phase.

\section{Quantum Phases in finite lattices with Periodic boundaries}

\label{sec-resultsfinite}

We now discuss results that demonstrate the competition between various Mott
and superfluid phases in the presence of SOC. We first present our results
on small system sizes with periodic boundaries. These system sizes are
consistent with small states formed in the center of traps in experiments.

Fig.~\ref{fig-phase} shows the phase diagram for four different limits of
the model, Eq.~(\ref{eq-totalmodel}). Fig.~\ref{fig-phase}a plots the
Bose-Hubbard phase diagram \cite{Fisher89} that results from setting the SOC
term and the inter-spin interaction term to zero in Eq.~(\ref{eq-totalmodel}%
), i.e., $\lambda=U_{\uparrow\downarrow}=0$. The absence of inter-spin
interactions allows two identical copies of the Mott insulator. The lower
and upper Mott lobes in Fig.~\ref{fig-phase}a correspond to $\langle
n_{i\uparrow} \rangle=\langle n_{i\downarrow}\rangle=1$ and $\langle
n_{i\uparrow}\rangle= \langle n_{i\downarrow} \rangle=2$, respectively.

Fig.~\ref{fig-phase}c shows the result of adding inter-spin repulsion, $%
U_{\uparrow\downarrow}>0$, but with no SOC, $\lambda=0$. Here we see that
that the original low energy Mott lobe is pushed up. The appearance of the
small Mott lobes (above and below the larger Mott lobe) correspond to the
formation of Mott insulators with Ising-type spin ordering. To see this, we
rewrite the interaction terms in $H$ using sum and difference operators, $%
n_{i\pm}\equiv n_{i\uparrow}\pm n_{i\downarrow} $. The large Mott lobe in
Fig.~\ref{fig-phase}c then corresponds to $\langle n_{i+}\rangle =2, \langle
n_{i-}\rangle =0$. The upper and lower small Mott lobes exhibit degeneracies
(for $t=0$) and correspond to $\langle n_{i+}\rangle =3, \langle
n_{i-}\rangle =\pm 1$ and $\langle n_{i+}\rangle =1, \langle n_{i-}\rangle
=\pm 1$, respectively. Here we exclude super exchange effects, $\mathcal{O}%
(t^{2}/U)$, discussed in other work \cite{Radic12,Cole12,Cai12,Gong12}.

\begin{figure}[t]
\centering
\includegraphics[width=3.3in]{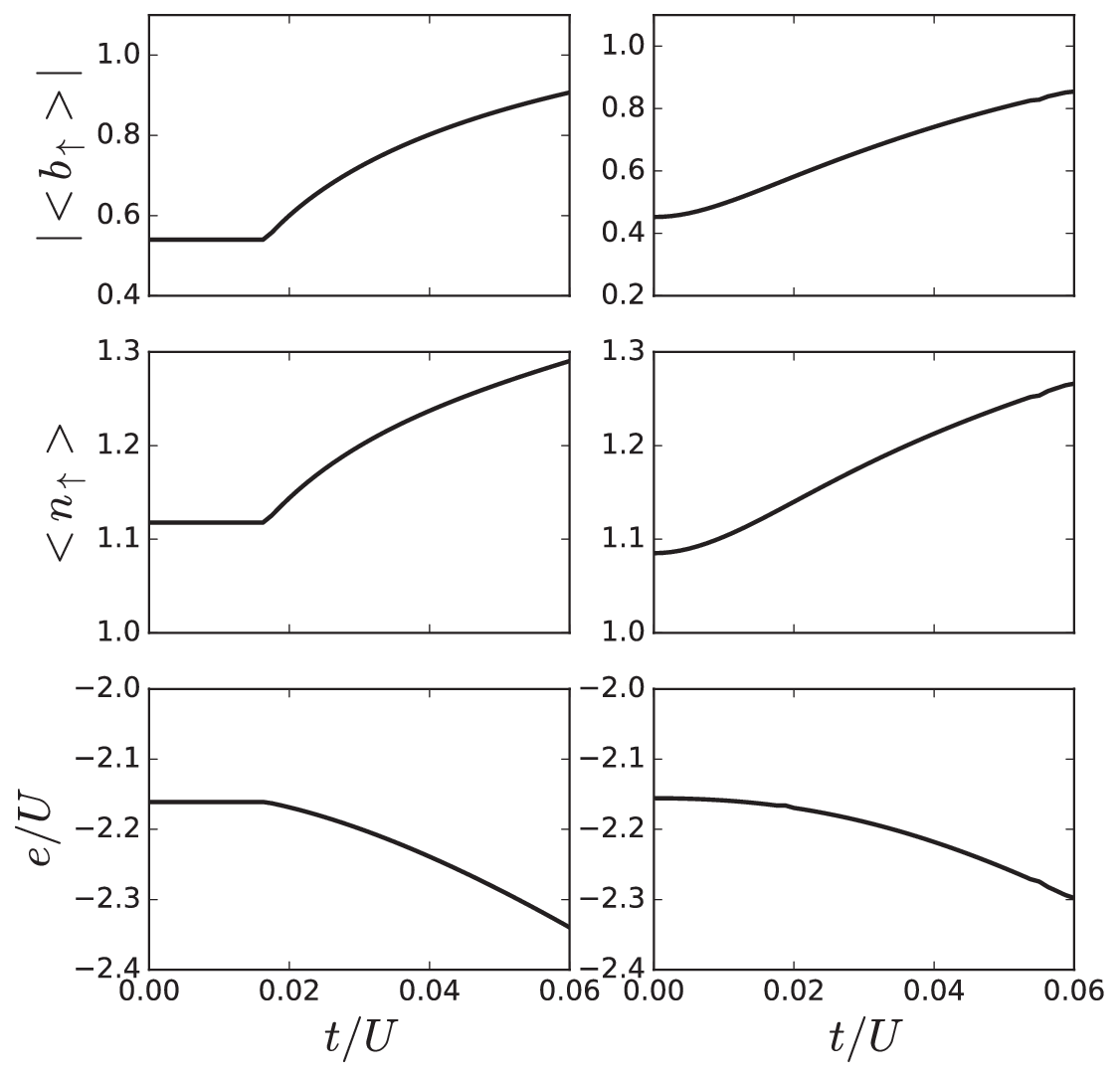}
\caption{Plot of the amplitude of spin-up superfluid order parameter $|
\langle b_{\uparrow} \rangle |$, the filling factor $\langle n_{\uparrow}
\rangle$ and the energy density $e$ as a function of the spin-independent
tunneling at $U_{\uparrow \downarrow} = 0.5U$, $\protect\lambda = 0.04U$ and
$\protect\mu = 1.33U$ for periodic (left panel) and open (right panel)
boundary conditions.}
\label{fig-trans}
\end{figure}

We now discuss the phase diagram that results from adding SOC. Figs.~\ref%
{fig-phase}b and \ref{fig-phase}d plot the phase diagrams that result from
adding SOC to the states depicted in Figs.~\ref{fig-phase}a and \ref%
{fig-phase}c, respectively. In both figures we see that the Mott insulators
at higher $\mu$ vanish. Increasing $\mu$ causes a direct transition from a
Mott insulator to a SOC-generated superfluid. At $t=0$, SOC \emph{alone}
drives the formation of a superfluid. We find that the Mott insulators that
normally persist at $t=0$ for all $\mu$ are actually supplanted by
SOC-generated superfluids. The $t=0$ superfluids found on this part of the
phase diagram derive kinetics purely from the spin-dependent tunneling in
SOC. We therefore find that even in the limit of vanishing kinetics, the
Rashba effect drives the Mott insulator into a conducting state (in this
case, a superfluid state). We have also checked the phase diagrams of $4
\times 4$ and $6 \times 6$ lattices, and find no qualitvative difference
with an $8 \times 8$ lattice shown in Fig.~\ref{fig-phase}. Below we show
that the precise nature of the resulting superfluid depends on the relative
strengths of $\lambda$ and $t$, as well as boundary effects.
\begin{figure}[t]
\centering
\includegraphics[width=3.3in]{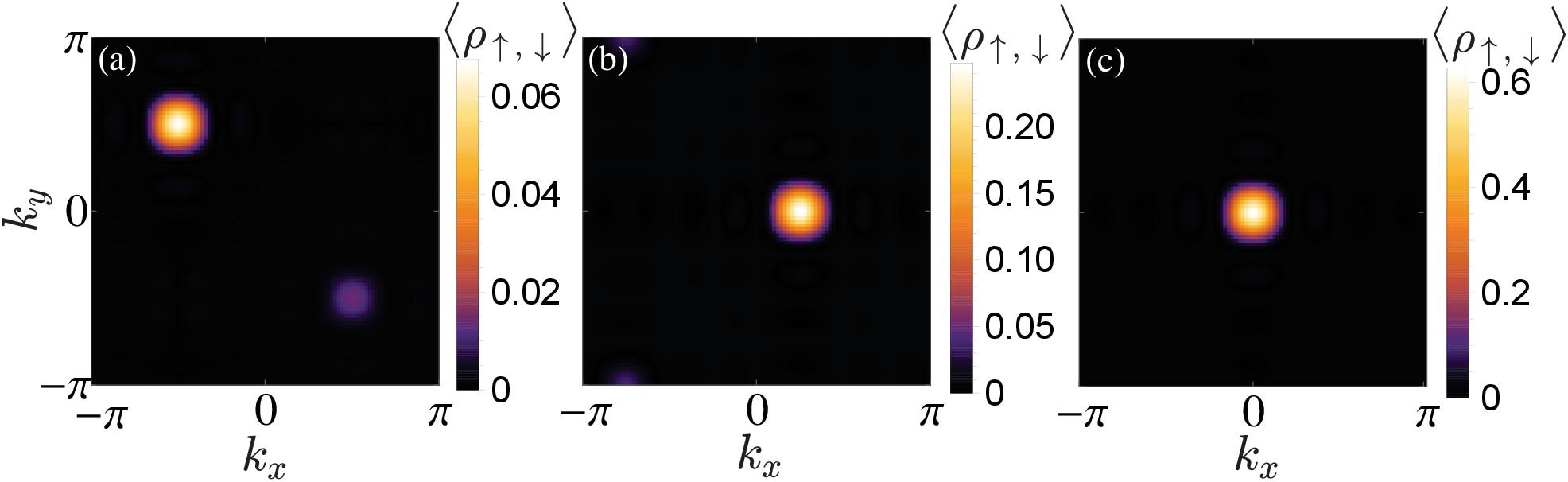}
\caption{(Color online) Spin-dependent momentum distribution, Eq.~(\protect
\ref{eq-sc}), for different superfluids at $U_{\uparrow \downarrow }=0.5U$, $%
\protect\lambda =0.04U$, $\protect\mu =1.33U$, (a) $t=0.005U$, and (b) $%
t=0.05U$.}
\label{fig-momentum}
\end{figure}

Fig.~\ref{fig-trans} shows the transitions of different superfluid phase
patterns. The left column shows the order parameters for the $8 \times 8$
lattice with periodic boundary conditions. Here SOC dominates and the
nonzero order parameters are unchanged for $t \le 0.019U$. For $t > 0.019U$,
the order parameter gradually increases with $t$, which indicates a
transition between different superfluids at $t \sim 0.019U$. For the open
boundary condition case shown in the right column, there is no such
transition since the phase can vary smoothly over the lattice.

The superfluids with different phase patterns have different momenta. To see
this we compute the spin-dependent momentum distribution at wavevector $k$:
\begin{equation}
\langle \rho _{\uparrow ,\downarrow }(\Vec{k})\rangle
=N^{-2}\sum_{i,j}\langle b_{i\uparrow }^{\dagger }b_{j\downarrow }^{%
\vphantom{\dagger}}\rangle e^{i\Vec{k}\cdot \left( \Vec{R}_{i}-\Vec{R}%
_{j}\right) },  \label{eq-sc}
\end{equation}%
where the lattice spacing is chosen as the unit of distance and $\Vec{R}_{j}$
is the location of the lattice site $j$.

We take random initial guess states and minimize the total energy to compute
the ground state $|G\rangle$, with which the spin-dependent momentum
distribution is computed as $\langle G|\rho _{\uparrow ,\downarrow }(\Vec{k}%
)|G\rangle /\langle G|G\rangle $. We get four degenerate ground states with
different momentum distributions, where the $D_4$ symmetry of the lattice
system is spontaneously broken. Similar results have been discovered in the
continuum model of spin-$1/2$ Bose- Einstein condensate with Rashaba SOC
\cite{Zhai10, Ho11, Pitaevskii12}, where the ground state is a single
plane-wave state with finite momentum, and the direction of plane wave is
spontaneously determined when the inter-spin interaction is smaller than the
intra-spin interaction.

The two different states in Fig.~\ref{fig-momentum} have qualitatively
distinct momentum distributions. We have also verified that in these two
phases, the magnitude of the order parameter is uniform over the whole
lattice, indicating that only the phase pattern changes during the
transition. Note that the peak in the momentum distribution for the first
two phases depends strongly on the ratio between $\lambda $ and $t$. In the
non-interacting limit, the ground state energy of the system with SOC is $%
E=-2t(\cos{k_x}+\cos{k_y})-2\lambda \sqrt{\sin^2{k_x}+\sin^2{k_y}}.$ The
energy minima are located at $\mathbf{k}=\big(\pm \arctan(\lambda/\sqrt{2}%
t),\pm \arctan(\lambda/\sqrt{2}t)\big).$ On a finite $8 \times 8$ lattice, $%
\mathbf{k}$ can only take discrete values. In Particular, for $\lambda/t=0.8$%
, the energy minima are located at $(0,\pi/4)$, $(0,-\pi/4)$, $(\pi/4,0)$
and $(-\pi/4,0).$ In the presence of interactions, $D_4$ symmetry is
spontaneously broken and the system chooses one of the minima in Fig. 5(b).
Similarly, for $\lambda/t=8$, the the energy minima are $\mathbf{k}=(\pm
\pi/2,\pm\pi/2)$, which is consistant with Fig. 5(a). It is therefore
possible to directly infer their ratio from the position of the peaks. We
also note that the results presented in Fig.~\ref{fig-momentum} relate
directly to the time-of-flight imaging that can measure momentum
distribution of distinct hyperfine states.

\section{Quantum phases in a Trapping Potential}

\label{sec-resultstrap}

\begin{figure}[t]
\centering
\includegraphics[width=3.3in]{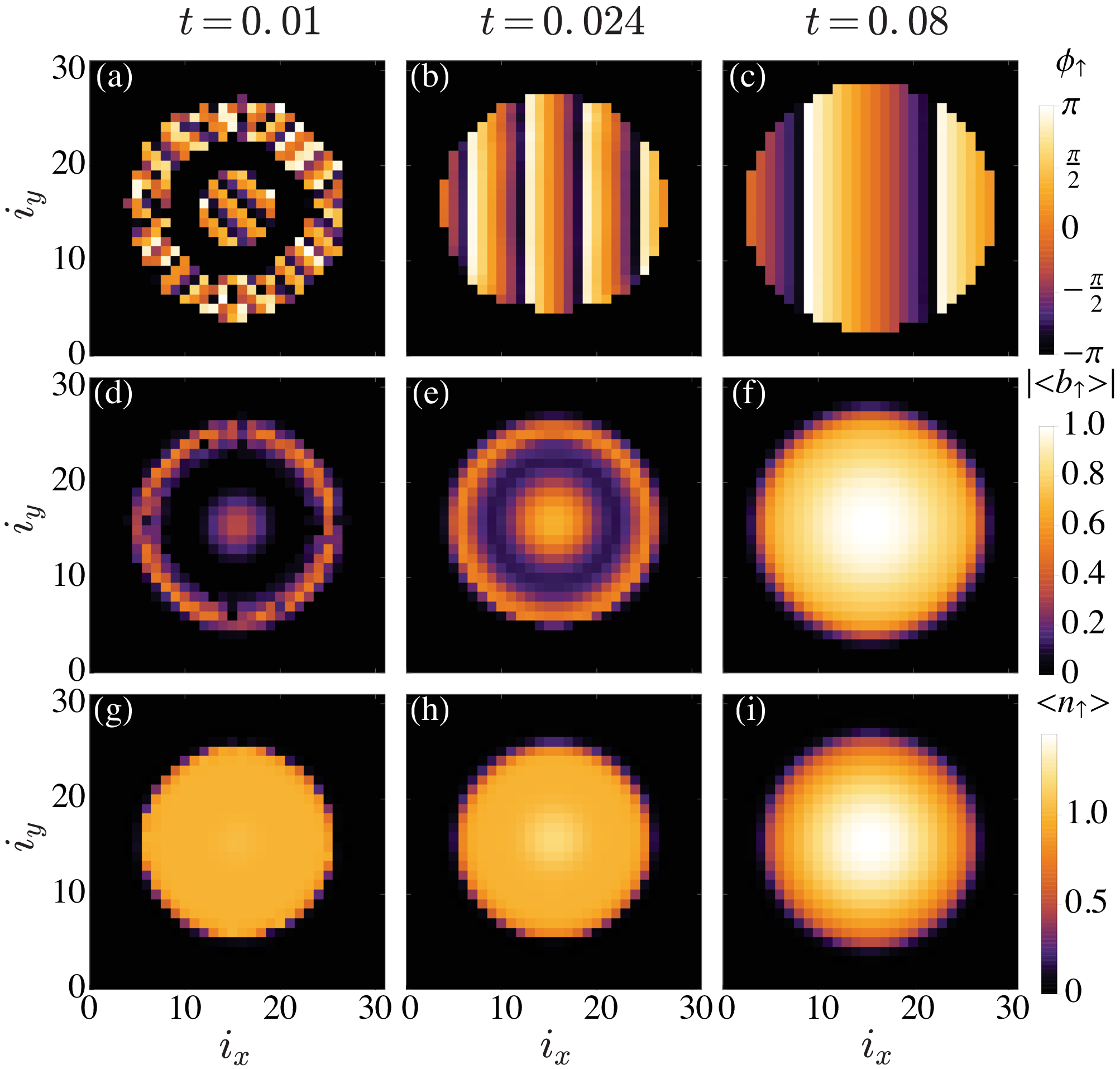}
\caption{Correlation functions of finite momentum superfluids on a $32
\times 32$ lattice with a confining potential [Eq.~\protect\ref%
{eq_confinement}] for $\protect\mu =0.8U$, $U_{\uparrow \downarrow} = 0$ and
$\protect\lambda =0.04U$. The left column shows results for $t=0.01U$ , the
middle column for $t=0.024U$ and the right column for $t=0.08U$. The top
three panels plot the phase $\protect\phi_{\uparrow}$ of the spin up
superfluid order parameter. The middle three panels plot the magnitude and
the bottom three panels plot the density. The phase patterns in the top two
panels reveal a sudden change in superfluid order.}
\label{fig-large}
\end{figure}

We now consider the effects of realistic confinement on the superfluid
transitions. The finite momentum superfluids considered here are akin to the
FFLO phase discussed in the context of trapped atomic Fermi gases. The FFLO
state depends strongly on lattice geometry. Finite size effects are normally
not considered to be relevant in solids because system sizes are typically
much larger than correlation lengths. But cold atomic gases can be put into
regimes where the system size is on the order of superfluid correlation
lengths.

Small magneto-optical trapping potentials can be created in cold atom
systems. We add a spatially varying chemical potential term to Eq.~(\ref%
{eq-totalmodel}) to model confinement: $\sum_{i}V(\Vec{R}_{i})(n_{i,%
\uparrow}+n_{i,\downarrow})$. The trapping potentials are well approximated
by a parabolic potential. We consider:
\begin{equation}
V(\Vec{R}_{i}) = 0.008U \left[\left(R_{i}^{x}-\frac{L_x-1}{2}\right)^2 +
\left(R_{i}^{y}-\frac{L_y-1}{2}\right)^2\right]  \label{eq_confinement}
\end{equation}
where $R_{i}^{x} (R_{i}^{y})$ is the $x(y)$ coordinate of site $i$ and $L_x$
($L_y$) is the lattice size along the $x$ ($y$) direction. The trap
coefficient is chosen to ensure that the trapped atom density vanishes
before the edge of the lattice is reached. Within the mean-feild theory, we
can compute the local superluid order parameter in the trap$\langle
b_{i,\sigma} \rangle = \sum_{n}\sqrt{n}f^{(i,\sigma)
*}_{n-1}f^{(i,\sigma)}_{n}.$ The local density is obtainted as $\langle
n_{i,\sigma} \rangle = \sum_{n}n|f^{(i,\sigma)}_{n}|^2.$

We now show that the phase change, discussed in periodic systems above, also
manifests in trapped systems. Fig.~\ref{fig-large} shows a typical example
obtained from solving Eq.~(\ref{eq-totalmodel}) in the presence of parabolic
trapping using the Gutzwiller ansatz with $10^4$ random initial guess
states. Since Mott insulator is a incoherent sate with random pahses, phases
of uparrow superfluid order parameter with $|\langle  b_{\uparrow } \rangle
| \le 0.05 $ are plotted with dark grey color in the top panel of Fig.~\ref%
{fig-large}. As the hopping parameter increases, the phase reorients in the
trap from non-uniform pattern to uniform due to the SOC effect. The effects
predicted here are observable in measurements sensitive to the phase of the
superfluid order parameter (e.g., the momentum distribution function). This
calculation shows that realistic trapping potentials lead to finite sized
systems that harbor the transitions found in periodic systems discussed
above.

\section{Quantum phases in Infinite Lattices}

\label{sec-resultsinfinite}

\begin{figure}[t]
\centering
\includegraphics[width=3.3in]{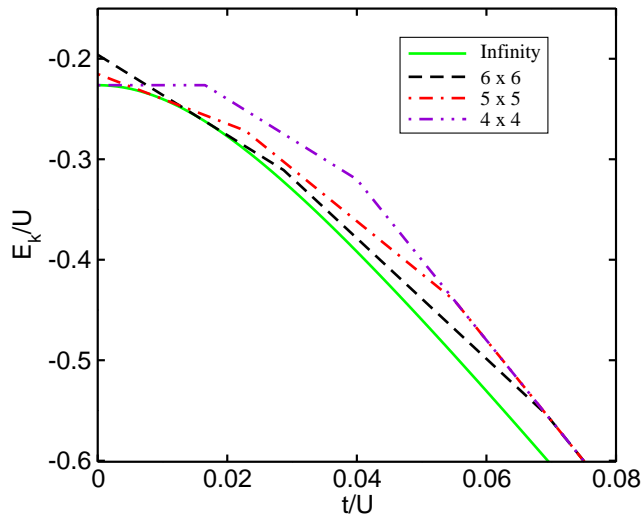}
\caption{Plot of the kinetic energy terms, Eq.~(\protect\ref{eq-eprime}), as
a function of the spin-independent tunneling at $\protect\lambda = 0.04U$,
for a $4 \times 4$, $5\times 5$, $6\times 6$ lattice and an infinite-size
lattice.}
\label{fig-analy}
\end{figure}

So far our study has been limited to finite-sized lattices. Here boundary
effects put a strong constraint on the superfluid phase patterns that can be
realized. But we can use Eq.~(\ref{eq-energymeanfield}) to study infinite
lattices. We find a general solution for the lowest-energy state, $\theta
_{i\uparrow }=\alpha (R_{i}^{y}-R_{i}^{x})$ and $\quad \theta _{i\downarrow
}=\frac{\pi }{4}+\alpha (R_{i}^{y}-R_{i}^{x})$, where $\alpha =\arctan
(\lambda \sqrt{2}/2t)$. The corresponding energy for just the kinetic terms
is:
\begin{equation}
E_{k}=-(tA+\lambda B)\psi ^{2}  \label{eq-eprime}
\end{equation}%
The competing superfluids arise from the competition between $A$ and $B$
coefficients.

Before studying the infinite system case we first test that Eq.~(\ref%
{eq-eprime}) gives the same results as the Gutzwiller mean field theory. We
find that this is the case by comparing results obtained from maximizing $tA
+ \lambda B$ in Eq.~(\ref{eq-eprime}) on a finite lattice with the
Gutzwiller mean field theory. We find precisely the same phase patterns
given in Fig.~\ref{fig-momentum}. This confirms that the Gutzwiller mean
field theory is equivalent to Eq.~(\ref{eq-eprime}) on finite lattices.

We now study infinite lattice sizes. In the infinite system size limit we
find: $E_k\rightarrow - 4\sqrt{2}\lambda \sin(\alpha) - 8t\cos(\alpha)$.
This implies that the energy will change smoothly as the period of the
finite momentum superfluids changes dramatically. Fig.~\ref{fig-analy} shows
that the energy computed on the infinite system size is in fact smooth. We
therefore conclude that infinite lattice sizes will eliminate transitions
observed in finite sized systems. A similar result was found in studies of
FFLO superfluids where periodic boundaries also constrain the FFLO momentum
to select certain values \cite{Samokhvalov10,Devreese11}. But we note that
realistic experiments are actually trapped finite sized systems with $N\sim
10^{2}-10^{5}$. We therefore conclude that transitions between distinct
superfluids found here should be observable in the small system limit
defined by the trap center.

\section{Summary}

\label{sec-summary}

We have studied the interplay of strong interaction and Rashba SOC in a
model motivated by optical lattice experiments: a 2D Hubbard model of
two-component bosons. We used mean field theory to map out the phase diagram
and study transitions. We find that strong Rashba SOC can completely destroy
the Mott insulator state, even in the absence of spin-independent tunneling
in the lattice. The Rashba SOC leads to superfluids with complex phase
patterns and finite momentum. We identified transitions between superfluids
with two different staggered phase patterns, that can be identified in the
spin-dependent momentum distribution. The spin-dependent momentum could be
accessed in time-of-flight measurements on optical lattices. We expect these
transitions to occur in finite sized systems but the phase patterns and
precise momenta depend strongly on the boundaries. We checked that these
transitions in phase patterns become smooth in infinite system sizes.

Our work relates to the nature of Mott insulator states in solids. Our study
of a 2D lattice finds that it is in principle possible for strong Rashba SOC
to convert a Mott insulator into a conducting state even in the limit of
vanishing kinetics ($t\rightarrow 0$ with $\lambda\sim 1$ in the lattice
model or $\langle p \rangle \rightarrow 0$ with $\langle\Vec{\nabla} V\times
\Vec{p} \rangle \sim E_F$ in the continuum). This limit could have bearing
on the nature of 3D Mott insulator surface states that experience very weak
kinetics but strong electric fields.

\emph{Acknowledgements} Y. Q., V. S., and C. Z. are supported by ARO
(W911NF-12-1-0335, W911NF-16-1-0182,W911NF-17-1-0128), AFOSR
(FA9550-11-1-0313,FA9550-15-1-0445, FA9550-16-1-0387), NSF-PHY (1505496),
and DARPA-YFA. M.G is supported in part by Hong Kong RGC/GRF Project 401512,
the Hong Kong Scholars Program (Grant No. XJ2011027) and the Hong Kong GRF
Project 2130352.


\begin{thebibliography}{99}
\bibitem{rashba:1960} E. I. Rashba, Sov. Phys. Solid State \textbf{2}, 1109
(1960).

\bibitem{AgAu} G. Nicolay, F. Reinert, S. Hufner, and P. Blaha, Phys. Rev. B
\textbf{65}, 033407 (2001).

\bibitem{AgAu1} G. Bihlmayer, S. Blugel, and E. V. Chulkov, Phys. Rev. B
\textbf{75}, 195414 (2007).

\bibitem{winkler:2003} R. Winkler, Spin-Orbit Coupling Effects in
Two-Dimensional Electron and Hole Systems, Springer Tracts in Modern Physics
(Springer, Berlin, 2003).

\bibitem{hasan:2010} M. Z. Hasan and C. L. Kane, Rev. Mod. Phys. \textbf{81}%
, 3045 (2010).

\bibitem{qi:2011} X. Qi and S. C. Zhang, Rev. Mod. Phys. \textbf{83}, 1057
(2011).

\bibitem{lin:2011} Y.-J. Lin, K. Jimenez-Garcia, and I. B. Spielman, Nature
\textbf{471}, 83 (2011).

\bibitem{wang:2012} P. Wang, Z. Q. Yu, Z. Fu, J. Miao, L. Huang, S. Chai, H.
Zhai, and J. Zhang, Phys. Rev. Lett. \textbf{109}, 095301 (2012).

\bibitem{cheuk:2012} L. Cheuk, A. T. Sommer, Z. Hadzibabic, T. Yefsah, W. S.
Bakr, and M. W. Zwierlein, Phys. Rev. Lett. \textbf{109}, 095302 (2012).

\bibitem{zhang:2012} J. Y. Zhang, S. C. Ji, Z. Chen, L. Zhang, Z. D. Du, B.
Yan, G. S. Pan, B. Zhao, Y. J. Deng, H. Zhai, S. Chen, and J. W. Pan, Phys.
Rev. Lett. \textbf{109}, 115301 (2012).

\bibitem{fu:2013} Z. Fu, L. Huang, Z. Meng, P. Wang, X. J. Liu, H. Pu, H.
Hu, and J. Zhang, Phys. Rev. A \textbf{87}, 053619 (2013).

\bibitem{Qu:2013} C. Qu, C. Hamner, M. Gong, C. Zhang, P. Engels, Phys. Rev.
A \textbf{88}, 021604(R) (2013).

\bibitem{williams:2013} R. A. Williams, M. C. Beeler, L. J. LeBlanc, K.
Jimenez-Garcia and I. B. Spielman, Phys. Rev. Lett. \textbf{111}, 095301
(2013).

\bibitem{bloch:2008} I. Bloch, J. Dalibard, and W. Zwerger, Rev. Mod. Phys.
\textbf{80}, 885 (2008).

\bibitem{galitski:2013} V. Galitski and I. B. Spielman, Nature \textbf{494},
49 (2013).

\bibitem{sau:2011} J. D. Sau, R. Sensarma, S. Powell, I. B. Spielman, and S.
Das Sarma, Phys. Rev. B \textbf{83}, 140510(R) (2011).

\bibitem{zhang:2013} Y. Zhang and C. Zhang, Phys. Rev. A \textbf{87}, 023611
(2013).

\bibitem{lin:2013} F. Lin, C. Zhang, and V. W. Scarola, Phys. Rev. Lett.
\textbf{112}, 110404 (2014).

\bibitem{hui:2017} H.-Y. Hui, Y. Zhang, C. Zhang, and V. W. Scarola, Phys.
Rev. A \textbf{95}, 033603 (2017).

\bibitem{chen:2017} M. Chen and V. W. Scarola, Phys. Rev. A \textbf{94},
043601 (2016).

\bibitem{hu:2012} H. Hu, B. Ramachandhran, H. Pu, and X. Liu, Phys. Rev.
Lett. \textbf{108}, 010402 (2012); B. Ramachandhran, H. Hu, and H. Pu, Phys.
Rev. A \textbf{87}, 033627 (2013).

\bibitem{zhou:2013} C. Wu, I. Mondragon-Shem, and X. Zhou, Chin. Phys. Lett.
\textbf{28}, 097102 (2011); X. Zhou, Y. Li, Z. Cai, and C. Wu, J. Phys. B:
At. Mol. Opt. Phys. \textbf{46}, 134001 (2013).

\bibitem{ramachandhran:2012} B. Ramachandhran, B. Opanchuk, X.-J. Liu, H.
Pu, P. D. Drummond, and H. Hu, Phys. Rev. A \textbf{85}, 023606 (2012).

\bibitem{Radic12} J. Radic, A. D. Ciolo, K. Sun, V. Galitski, Phys. Rev.
Lett. \textbf{109}, 085303 (2012).

\bibitem{Cole12} W. S. Cole, S. Z. Zhang, A. Paramekanti, and N Trivedi,
Phys. Rev. Lett. \textbf{109}, 085302 (2012).

\bibitem{Cai12} Z. Cai, X. Zhou, C. Wu, Phys. Rev. A \textbf{85}, 061605(R)
(2012).

\bibitem{Mandal} Saptarshi Mandal, Kush Saha, K. Sengupta, Phys. Rev. B
\textbf{86}, 155101, (2012).

\bibitem{Gong12} M. Gong, Y. Qian, V. W. Scarola, C. Zhang, arXiv:1205.6211.

\bibitem{Band} I. Vurgaftman, J. R. Meyer, and L. R. Ram-Mohan, J. Appl.
Phys. \textbf{89}, 5815 (2001).

\bibitem{agaas} J. B. Miller, D. M. Zumbuhl, C. M. Marcus, Y. B.
Lyanda-Geller, D. Goldhaber-Gordon, K. Campman, and A. C. Gossard, Phys.
Rev. Lett. \textbf{90}, 076807 (2003).

\bibitem{ggaas} M. Oestreich and W. W. Ruhle, Phys. Rev. Lett. \textbf{74},
2315 (1995).

\bibitem{ainas} G. L. Chen, J. Han, T. T. Huang, S. Datta, and D. B. Janes,
Phys. Rev. B \textbf{47}, 4084 (1993).

\bibitem{Lg} H. Kosaka, Electronics Letters \textbf{37}, 464 (2001).

\bibitem{ainsb} S. K. Greene, J. Singleton, P. Sobkowicz, T. D. Golding, M.
Pepper, J. Perenboom, and J. Dinan, Semicond. Sci. Technol. \textbf{7}, 1377
(1992).

\bibitem{Patrick} K. Michaeli, A. C. Potter, and P. A. Lee, Phys. Rev. Lett.
\textbf{108}, 117003 (2012).

\bibitem{FF64} P. Fulde, and R.A. Ferrell, Phys. Rev. \textbf{135}, A550
(1964). 

\bibitem{LO64} A.I. Larkin, and Yu.N. Ovchinnikov, Zh. Eksp. Teor. Fiz.
\textbf{47}, 1136 (1964). 

\bibitem{LO65} A.I. Larkin, and Yu,N. Ovchinnikov, Sov. Phys. JETP \textbf{20%
}, 762 (1965). 

\bibitem{FFLO:Wu} Z. Cai, Y. Wang, and C. Wu, Phys. Rev. A \textbf{83},
063621 (2011).

\bibitem{RMP04} R. Casalbuoni and G. Nardulli, Rev. Mod. Phys. \textbf{76},
263 (2004).

\bibitem{Mizushima05} T. Mizushima, M. Ichioka, K. Machida, Journal of
Physics and Chemistry of Solids \textbf{66}, 1359 (2005).

\bibitem{Hu07} H. Hu, X. Liu, and P. D. Drummond, Phys. Rev. Lett. \textbf{98%
}, 070403 (2007).

\bibitem{Koponen08} T. K. Koponen, T. Paananen, J.-P. Martikainen, M. R.
Bakhtiari and P. T\"{o}rm\"{a}, New J. Phys. \textbf{10}, 045014 (2008).%

\bibitem{Feiguin07} A. E. Feiguin and F. Heidrich-Meisner, Phys. Rev. B
\textbf{76}, 220508(R) (2007).%

\bibitem{Paananen09} T. Paananen, J. Phys. B: At. Mol. Opt. Phys. \textbf{42}%
, 165304 (2009).

\bibitem{Parish07} M. M. Parish, S. K. Baur, E. J. Mueller, and D. A. Huse,
Phys. Rev. Lett. \textbf{99}, 250403 (2007).

\bibitem{Tempere07} J. Tempere, M. Wouters, and J. T. Devreese, Phys. Rev. B
\textbf{75}, 184526 (2007).

\bibitem{Loh10} Y. L. Loh and N. Trivedi, Phys. Rev. Lett. \textbf{104},
165302 (2010).

\bibitem{Jaksch98} D. Jaksch, C. Bruder, J. I. Cirac, C. W. Gardiner, and P.
Zoller, Phys. Rev. Lett. \textbf{81}, 3108 (1998).

\bibitem{rokhsar:1991} D.S. Rokhsar and B.G. Kotliar, Phys. Rev. B \textbf{44%
}, 10328 (1991).

\bibitem{scarola:2007} V. W. Scarola and S. Das Sarma, Phys. Rev. Lett .
\textbf{98}, 210403 (2007).

\bibitem{zakrezewski:2005} J. Zakrzewski, Phys. Rev. A \textbf{71}, 043601
(2005).

\bibitem{niederle:2013} A. Niederle and H. Rieger, New J. Phys. \textbf{15},
075029 (2013).

\bibitem{Debye09} P. Debye, Mathematische Annalen \textbf{67}, 535 (1909).

\bibitem{Deift93} P. Deift, X. Zhou, Ann. Of Math. \textbf{137}, 295 (1993).

\bibitem{Fisher89} M. P. A. Fisher, P. B. Weichman, G. Grinstein, and D. S.
Fisher, Phys. Rev. B \textbf{40}, 546 (1989).

\bibitem{Zhai10} C. J. Wang, C. Gao, C. M. Jian, and H. Zhai, Phys. Rev.
Lett. 105, 160403 (2010).

\bibitem{Ho11} T.-L. Ho and S. Zhang, Phys. Rev. Lett. 107, 150403 (2011)

\bibitem{Pitaevskii12} Y. Li, L. P. Pitaevskii, and S. Stringari, Phys. Rev.
Lett. 108, 225301 (2012)

\bibitem{Samokhvalov10} A. V. Samokhvalov, A. S. Melnikov, and A. I. Buzdin,
Phys. Rev. B \textbf{82}, 174514 (2010).

\bibitem{Devreese11} J. P. A. Devreese, M. Wouters, and J. Tempere, Phys.
Rev. A \textbf{84}, 043623 (2011).
\end{thebibliography}
\end{document}